\documentstyle[twoside,fleqn,espcrc2]{article}

\newcommand{\AmS}{{\protect\the\textfont2
  A\kern-.1667em\lower.5ex\hbox{M}\kern-.125emS}}

\title{Algorithmic aspects of multicanonical simulations}

\author{Bernd A. Berg\address{Department of Physics and
Supercomputer Computations Research Institute, \\
Florida State University, Tallahassee, FL 32306, USA} }
       
\begin{document}

\begin{abstract}
Monte Carlo (MC) simulations of many systems, in particular those 
with conflicting constraints, can be considerably speeded up by using 
multicanonical or related methods. Some of these approaches
sample with a-priori unknown weight factors. After introducing
the concept, I shall focus on two aspects: (i) Opinions about the 
optimal choice of weight factors. (ii) Methods to get weight factor
estimates, with emphasize on a multicanonical recursion.
\end{abstract}

\maketitle

\section{INTRODUCTION}

One of the questions which ought to be addressed before performing
large scale computer simulations is ``What are suitable weight factors
for the problem at hand?''. It has been expert wisdom 
for quite a while \cite{ToVa77},
and became widely recognized in recent years [2-8], that MC simulations
with a-priori unknown weight factors are feasible and deserve to be
considered. 

The design of suitable weights requires to understand or guess relevant 
physical features of the system(s) under (numerical) investigation. 
This is already exemplified by the work of Metropolis et al.\cite{Me53}.
To sample the important configurations of a canonical ensemble at
temperature $T=\beta^{-1}$ efficiently, the Boltzmann weight is 
recommended

\begin{equation}
w(a) = e^{-\beta E_a}\, ,
\end{equation}
where $E_a$ is the energy of configuration $a$. Namely, it is 
straightforward to show that this weight minimizes the sampling error 
for the canonical (normalized) energy density $P(E)$:

\begin{equation}
\sum (\triangle P(E))^2 = {\rm minimum}\, .
\end{equation}
However, even when studying canonical statistical mechanics the 
Boltzmann weight is not a satisfactory recipe for all cases. 
It can be false that the weight factor (1) samples the important 
configurations. A well-studied case is the sampling of configurations
with interfaces \cite{BeNe92,Ja94}. Quite generally, instead of (2)
the sampling error for the actually calculated observable needs to be
optimized. 

A second reason, called ``dynamic'' in \cite{Be97}, relies on 
the fact that the error estimate (2) holds for independent
configurations. To sample with a weight like (1), one has to rely on
a Markov chain. This introduces an autocorrelation time which
depends on the weights chosen. In this sense the argument leading 
from (1) to (2) is ``static'' \cite{Be97}.
It can happen that statically most efficient weights are rendered 
inefficient by the dynamics, whereas statically worse weights can yield 
a favorable dynamics. For instance, in systems with conflicting 
constraints encouraging experience has been made with weights 
\cite{BeCe92,HaOk93} and expanded ensembles 
\cite{Ly92,MaPa92,KeRe94,HuNe96} which allow to ``refresh'' the
system in the disordered phase. It seems worthwhile to study 
this concept also for less difficult situations, like 
second order phase transitions.

In the next section I consider the  question of optimal weights. 
The third (and major)
section focuses on how to get working estimates of the weights in the
first place. Brief conclusions are given in section~4.

\section{WHICH WEIGHTS?}

First, I consider the static aspects. For the relative sampling error of
the
spectral density $n(E)$ we get

\begin{equation}
\sum (\triangle n(E) / n(E) )^2 = {\rm minimum}\, 
\end{equation}
by using 

\begin{equation}
w(a)  = 1/n(E_a)\, . 
\end{equation}
The result is an uniform energy probability density.
For a large number of applications this has proven to be a  robust choice,
for instance when dealing with first order phase transitions
\cite{BeNe92,Ja94}.  Instead of the energy, weights in other thermodynamic
variables have also been used successfully, for example $w(a)=f(m_a)$ 
\cite{BeHaNe93}, where $m_a$ is the magnetization of configuration $a$, 
or $w(a)=f(b_a)$ \cite{JaKa95}, where $b$ the bound variable for cluster
updating.

In some applications one is tempted to introduce weight factors in
two variables,  like $w(a)=f(E_a,m_a)$, which would allow to re-construct 
canonical expectation values over both, a macroscopic temperature and 
magnetic field range.  Only for exceptional cases, when one is interested
in 
very small volumes, this seems to be practical. In most cases the interest
lies
in a finite size scaling (FSS) analysis of the infinite volume limit
$V\to\infty$. 
CPU time slows down worse than $V^2$ for the discussed methods. One 
cannot afford to replace $V$ by $V^2$, i.e. a performance worse than $V^4$.
In addition, a RAM increase with $V^2$ is unpleasant too. The art seems
to be to find {\it one} combination of variables, 
such that suitable weighting with it leads to significant 
improvements of the numerical performance.

With respect to groundstate searches it is claimed in ref.\cite{HeSt95}
that the weight 

\begin{equation}
w(a) = 1 / \sum_{E'=E_{\min}}^{E_a} n(E')
\end{equation}
gives the best worst case performance in terms of ergodicity and
pertinence. 
The authors define pertinence by the static
argument that $V N_s$ independent samples with the weight (5) are better
than $N_s$ independent rival samples. Ergodicity relates to the dynamics of
the Markov chain. The latter is system dependent and there is presently
little understanding of the processes driving its slowing down. Hence,
it seems  to me that their statement about ergodicity is unwarranted.

The idea of expanded ensembles \cite{Ly92,MaPa92}, particularly attractive
in their version of parallel tempering \cite{Te96,HuNe96}, tries to overcome
dynamical slowing down by introducing moves into additional dimensions. 
Whereas new weights may alter or by-pass barriers, expanded (canonical) 
ensembles can  only by-pass them, because  Boltzmann weights are used for 
the normal updates. A, yet untested, possibility would be to apply the 
parallel tempering idea to unconventional ensembles. This may allow to 
combine the advantages of both approaches. 

\section{RECURSIVE WEIGHT ESTIMATES}

Weight factors like (4) or (5) are a-priori unknown. This introduces some
additional complications, unknown to canonical simulations.
Before the simulation can be performed, a working estimate of the weight
factor has to be known, {\it i.e.} an
approximation for the practical purpose of performing the MC simulation. 
Deviations by factors around ten may well be acceptable, because the
accuracy of the simulation is hardly affected by this. The errors of
the simulation remains in any  case statistical, because the actually used 
weight factors are exactly known. On the other hand, deviations 
from the desired weights by factors of, say, order $10^{10}$ (well the
order of improvements) are certainly no longer acceptable. Such large
discrepancies would entirely obstruct the basic ideas.

In some cases, for instance first order phase transitions \cite{BeHaNe93},
FSS allows reasonable working estimates on the basis of
already simulated smaller systems.  
Also constraint simulations have
successfully been used \cite{BiNeBe93,CsFo95}. 
For other situations, like spin glasses \cite{BeCe92}, this does not work. 
In the remainder of this section I will 
first explain the difficulties of the most straightforward recursion.
Subsequently, I give an efficient solution, which essentially is a
simplified and improved presentation of an result of ref.\cite{Be96}.

Desired is a working estimate of the weight factor (4). 
Others can be treated similarly. It is advisable to start any recursion in
the disordered phase of a system, where  the system moves most freely
under MC updates.  Therefore, let us assume 
for the zeroth simulation. 

\begin{equation} \label{w0}
w^0  (a) = 1 \ {\rm for\ all}\ a\, .
\end{equation}
The most obvious recursion goes now 
as follows: Simulation $n$, $(n=0,1,2,...)$ is carried out with the 
estimate $w^n (a) $  and yields the histogram $H^n (E)$. Estimate
$n+1$ for the weight factors is then given by

\begin{equation} \label{naive}
w^{n+1} (a) = { w^n (a) \over H^n (E_a) }\, . 
\end{equation}
This recursion has a number of  difficulties

\begin{description}
\item{(i)} What to do with histogram entries $H^n (E) = 0$ or small?
\item{(ii)} Each $H^n (E)$ calculation starts with zero statistics. Assume,
someone has given us the exact result and we use it for $w^0 (a)$:
The next estimate $w^1 (a)$ will be worse.  A noisy left-over of $w^0 (a)$.
\item{(iii)} Initial weights $w(a)=1$ correspond to temperature 
infinity and are bad in the limit 
$E_a\to E_{\min}$, approached towards zero temperature.
\end{description}
To avoid (i), one may replace 

\begin{equation} \label{hatH}
H(E)\to {\hat H(E)} = \max\, [h_0,H(E)]\, ,
\end{equation}
where $h_0$  is a number $\le 1$. The recursion derived in the 
following overcomes (ii).

Let us first discuss the relationship \cite{BeNe91,Be96} of the weight 
factors with microcanonical temperature $\beta (E)$, because it turn out
to be advantageous to derive the final recursion in this quantity. 
It holds

\begin{equation}
w^{n+1} (a) = e^{-S(E_a)} = e^{-\beta(E_a)\, E_a + \alpha (E_a)}\, 
\end{equation}
where $S(E)$ is the microcanonical entropy and,  by definition,

$$ \beta (E) = {\partial S (E) \over \partial E}\, .$$
This determines the fugacity function $\alpha (E)$ up to an 
(irrelevant) additive constant. Consider, for example, the case
of a discrete minimal energy $\epsilon$. We may choose

\begin{equation} \label{betaE}
\beta (E) = \left[ S(E+\epsilon) - S(E) \right] / \epsilon
\end{equation}
and the identity $S(E) = \beta(E)\, E - \alpha (E)$ implies
$ S(E) - S(E-\epsilon) = \beta(E) E - \beta(E-\epsilon) 
  (E-\epsilon ) -
   \alpha(E) + \alpha(E-\epsilon)$. Inserting $\epsilon\, 
\beta(E-\epsilon) = S(E) - S(E-\epsilon)$ yields 

\begin{equation}
\alpha(E-\epsilon) = \alpha(E) + \left[ \beta(E-\epsilon) 
 - \beta (E) \right]\, E
\end{equation}
and $\alpha(E)$ is fixed by defining $\alpha(E_{\max})=0$.
In summary, once $\beta (E)$ is given, $\alpha (E)$ follows for free.
The starting condition (\ref{w0}) becomes (other $\beta^0 (E)$ choices 
are of course possible)

\begin{equation}
\beta^0 (E)= \alpha^0 (E) = 0\, .
\end{equation}
With $\hat{H}$ defined by (\ref{hatH}) we translate now eqn.(7) into 
an equation for $\beta (E)$. Subscripts $_0$ are used to indicate
that those quantities are not yet the final estimators from the
$n^{th}$ simulation. Let
$$ w^{n+1}_0(E)=e^{-S^{n+1}_0(E)}=c\, {w^n(E) \over \hat{H}^n(E)}\, ,$$
where the (otherwise irrelevant) constant $c$ is introduced to ensure 
that $S^{n+1}_0(E)$ can be an estimator of the microcanonical entropy. 
It follows $S^{n+1}_0(E) = - \ln c+S^n(E)+\ln \hat{H}^n(E)$. 
Inserting this relation into (\ref{betaE}) gives

\begin{equation} \label{beta0} \begin{array}{c}
\beta^{n+1}_0 (E) = \beta^n (E) + ~~~~~~~~~~~~~~~~~~~~~~~~~~~ \\
~[ \ln \hat{H}^n(E+\epsilon) - \ln \hat{H}^n(E) ] / \epsilon
\end{array} \end{equation}
As estimator of the variance follows
$$\sigma^2[\beta^{n+1}_0(E)]
={c'\over H^n(E+\epsilon)}+{c'\over H^n(E)}\, ,$$
where $c'$ is an unknown constant and we have re-introduced the 
original histograms to emphasize (and use) that the variance is 
infinite when there is zero statistics. The statistical weight for 
$\beta^{n+1}_0 (E)$ is inversely proportional to its variance. 
Choosing a convenient constant, we get

\begin{equation}
g^n_0 (E) = {H^n (E+\epsilon)\ H^n (E) \over
H^n (E+\epsilon) + H^n (E)}\, .
\end{equation}
Note that $g^n_0(E)=0$ for $H^n(E+\epsilon)=0$ or $H^n(E)=0$.
The $n^{th}$ simulation was carried out using $\beta^n (E)$. It is
now straightforward to combine $\beta_0^{n+1} (E)$ and 
$\beta^n (E)$ according to their respective statistical weights
into the desired estimator:

\begin{equation} \label{beta}
\beta^{n+1} (E) = \hat{g}^n (E)\,  \beta^n (E) +
                \hat{g}^n_0 (E)\, \beta^{n+1}_0 (E)\, ,
\end{equation}
where the normalized weights
$$ \hat{g}^n_0 (E) = {g^n_0(E) \over g^n(E) + \hat{g}^n_0 (E)}\
{\rm and}\ \hat{g}^n (E) = 1 - \hat{g}^n_0 (E) $$
are determined by the recursion 

\begin{equation}
g^{n+1} (E) = g^n(E) + g^n_0(E),\ g^0(E)=0\, .
\end{equation}
We can eliminate $\beta^{n+1}_0 (E)$ from eqn.(\ref{beta}) by
inserting its definition (\ref{beta0}) and get

\begin{equation} \label{recursion} \begin{array}{c}
 \beta^{n+1}(E) = \beta^n (E) + \hat{g}^n_0(E) \times 
 ~~~~~~~~~~~~~~~~~ \\ 
~[ \ln \hat{H}^n(E+\epsilon)-\ln \hat{H}^n(E)] / \epsilon
\end{array} \end{equation}
Notice that the sole purpose of using $\hat{H}$, instead of $H$ has 
become to avoid $\ln(0)$. The weight $\hat{g}^n_0(E)$ of such contributions
is zero. The major advantage of (\ref{recursion}), compared with (\ref{naive}), 
is that it keeps already assembled statistics. This solves (ii) and eases
(iii). Frequent iterations are allowed, implying increased stability and 
decreased CPU time consumption. In addition one may implement
a suitable $\beta (E)$ guess for $E\to E_{\min}$.

Finally, eqn.(\ref{recursion}) can be converted into a recursion for ratios 
of weight factor neighbors. We define

\begin{equation} 
R^n (E) = e^{\epsilon\, \beta^n (E)} = {w^n(E) \over w^n (E+\epsilon)}
\end{equation}
and get the recursion

\begin{equation}
R^{n+1} (E) = R^n (E)\, \left[ {\hat{H}^n (E+\epsilon) \over
              \hat{H}^n (E) } \right]^{\hat{g}^n_0(E)}\, .
\end{equation}

\section{CONCLUSIONS}

Sampling of broad energy and other unconventional distributions is
technically feasible. Statical slowing down can be overcome, as is
well established for first order phase transitions. Dynamical slowing
down is more difficult, but further algorithmic improvements (find
the relevant combination of variables!) appear to be possible.

{\bf Acknowledgement:} I would like to thank Wolfhard Janke for
useful discussions.

\end{document}